\pdfoutput=1

\documentclass[10pt,letterpaper]{article}
\usepackage{opex3}
\usepackage{graphicx}

\begin{document}

\title{A metamaterial absorber for the terahertz regime: Design, fabrication and characterization}

\author{Hu Tao$^{1\dag}$, Nathan I. Landy$^{2\dag}$, Christopher M. Bingham$^2$, Xin Zhang$^1$, Richard D. Averitt$^3$, and Willie J. Padilla$^2$}
\address{
$^1$Boston University, Department of Manufacturing Engineering, 15 Saint Mary's Street, Brookline, Massachusetts 02446, USA.\\
$^2$Boston College, Department of Physics, 140 Commonwealth Ave., Chestnut Hill, MA 02467 USA.\\
$^3$Boston University, Department of Physics, 590 Commonwealth Avenue, Boston, Massachusetts 02215, USA.\\
$^{\dag}$ Contributed equally to this work.}

\email{Willie.Padilla@bc.edu} 



\begin{abstract}
We present a metamaterial that acts as a strongly resonant absorber
at terahertz frequencies. Our design consists of a bilayer unit cell
which allows for maximization of the absorption through independent
tuning of the electrical permittivity and magnetic permeability. An
experimental absorptivity of 70\% at 1.3 terahertz is demonstrated.
We utilize only a single unit cell in the propagation direction,
thus achieving an absorption coefficient $\alpha$ = 2000 cm$^{-1}$.
These metamaterials are promising candidates as absorbing elements
for thermally based THz imaging, due to their relatively low volume,
low density, and narrow band response.
\end{abstract}



\section{Introduction}

The electromagnetic response of natural materials forms the basis
for the construction of most modern optoelectronic devices. However,
this EM response is not evenly distributed across the
electromagnetic spectrum. At frequencies of a few hundred gigahertz
and lower, electrons are the principle particles which serve as the
workhorse of devices. On the other hand, at infrared through optical
/ UV wavelengths, the photon is the fundamental particle of choice.
In-between these two fundamental response regimes there exists a
region comparatively devoid of material response, commonly referred
to as the ``terahertz gap" (0.1-10 THz, $\lambda$=3mm-30$\mu$m)
\cite{williams06,tonouchi07}. Although enormous efforts have focused
on the search for ``terahertz" materials or alternative novel
techniques to enable the construction of device components, much
work remains. There is a wide range of natural phenomena that could
be probed with terahertz (THz) devices. Specifically, a THz detector
would be useful for imaging in areas such as biology
\cite{zhang02,crowe04} and security
\cite{oliveira03,zimdars03,federici05,liu06,barber05}.

\begin{figure}
[ptb]
\begin{center}
\includegraphics[keepaspectratio=true]%
{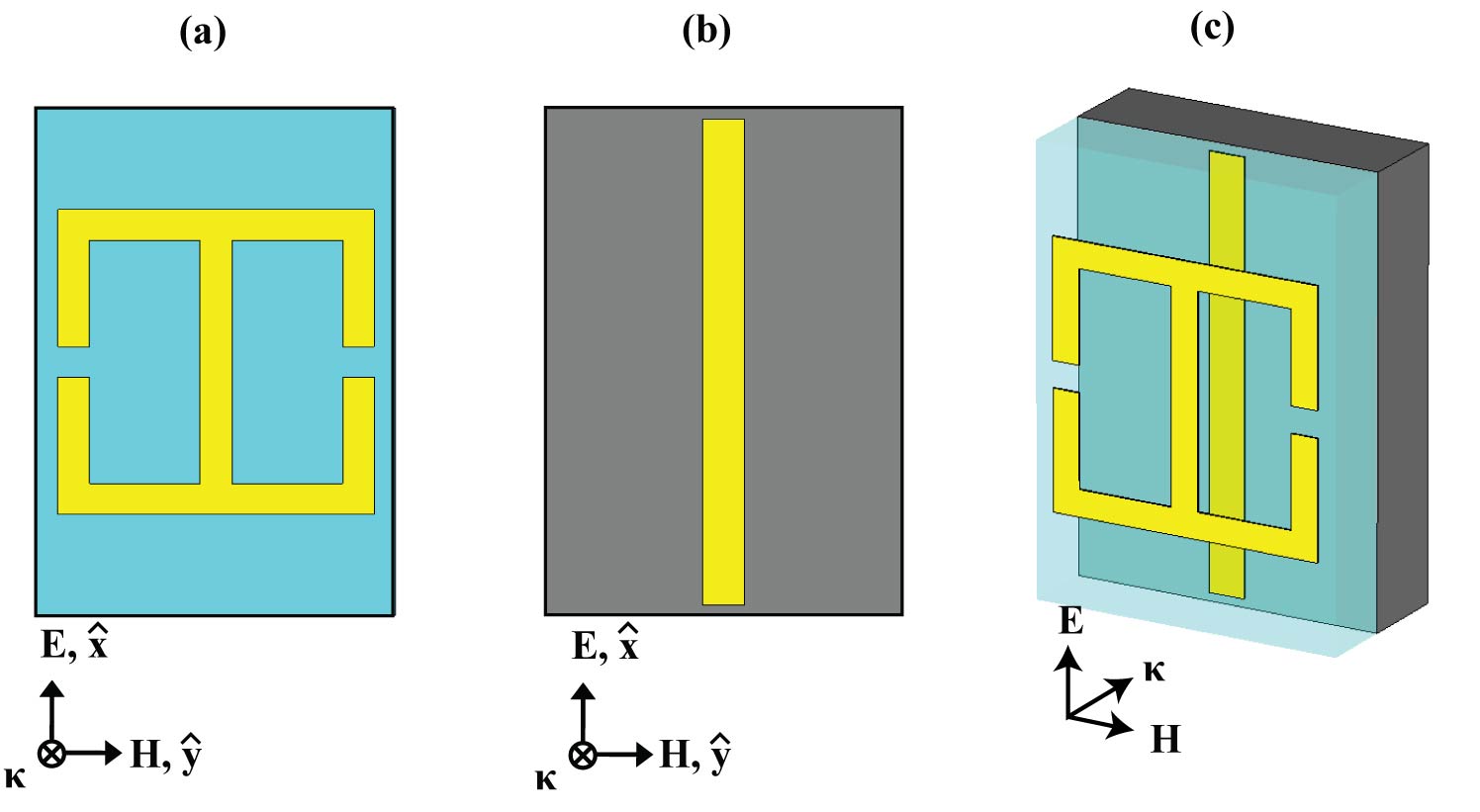}%
\caption{(color) Schematics of the THz absorber: a) electric
resonator on the top of a polyimide spacer; b) cut wire on GaAs
wafer; c) single unit cell showing the direction of propagation of
incident EM wave. The unit cell is 34 $\mu$m wide and 50 $\mu$m in
length. The line width and gap of the electric resonator is 3
$\mu$m. The side length of the square electric resonator is 30
$\mu$m, the side length of the cut wire is 48 $\mu$m, and the width
of the cut wire is 4 $\mu$m. Thickness of the electric resonant ring
and cut wire is 200 nm. The spacer of polyimide is 8 $\mu$m thick,
and the GaAs wafer is 500 $\mu$m thick.}
\label{Fig1}%
\end{center}
\end{figure}

Recently, there has been considerable effort to construct engineered
electromagnetic materials for operation specifically within the void
of natural material response described above
\cite{padilla07,yen04,chen06,padilla06}. These artificial systems,
called metamaterials, are composites whose EM properties originate
from oscillating electrons in unit cells comprised of highly
conductive and shaped metals such as gold or copper. The
sub-wavelength unit cell is replicated to form a material, which
allows for a designed resonant response of the metamaterial's
electrical and magnetic properties. Metamaterials can be regarded as
effective media and characterized by a complex electric permittivity
$\widetilde{\epsilon}(\omega) = \epsilon_1(\omega) +
i\epsilon_2(\omega)$ and complex magnetic permeability
$\widetilde{\mu}(\omega) = \mu_1(\omega) + i\mu_2(\omega)$. Resonant
structures that couple strongly to either the
electric~\cite{padilla07} or magnetic~\cite{yen04} fields have been
demonstrated at terahertz. Significant growth in metamaterial
research has been due to efforts to create negative refractive index
(NRI) materials \cite{smith00,shelby01,pendry00} and, more recently,
invisibility cloaks \cite{pendry06,schurig06a}. As such, the primary
focus has been on the index of refraction defined as
$\widetilde{n}(\omega)=
\sqrt{\widetilde{\epsilon}(\omega)\widetilde{\mu}(\omega)}$, where
one desires $n_1<0$ for negative index or $0<n_1<1$ for cloaks. To
create such structures, it is important to minimize losses over the
operating frequency range, which is associated with the imaginary
portion of the index, and thus strive for $n_2\rightarrow0$.
Conversely, for many other applications it would be desirable to
maximize the loss which is an aspect of metamaterial research that,
to date, has received very little attention. A recent example is the
creation of a resonant high absorber which has been demonstrated at
microwave frequencies \cite{landy07}. Such an absorber would be of
particular importance at terahertz frequencies where it is difficult
to find naturally occurring materials with strong absorption
coefficients that are also compatible with standard microfabrication
techniques. By fabricating bilayer metamaterial structures it
becomes possible to simultaneously tune
$\widetilde{\epsilon}(\omega)$ and $\widetilde{\mu}(\omega)$ such
that a high absorptivity can be achieved. In principle, this
tunability could lead to near unity absorptivity. In practice this
is limited by achievable fabrication tolerances.

We present a first generation terahertz metamaterial absorber which
achieves a resonant absorptivity of 70\% at 1.3 THz. Given the 8
$\mu$m thickness of our metamaterial, this corresponds to a power
absorption coefficient of $\alpha=$2000 cm$^{-1}$ which is
significant at THz frequencies. The strong absorption coefficient
makes this low volume structure a promising candidate for the
realization of enhanced, spectrally selective, thermal detectors. A
single unit of the absorber consists of two distinct metallic
elements: an electrical ring resonator Fig. \ref{Fig1} (a) and a
split wire Fig. \ref{Fig1} (b). The electrical ring resonator (ERR)
consists of two single split rings sitting back to back. The two
inductive loops are of opposite handedness and thus couple strongly
to a uniform electric field, and negligibly to magnetic
fields~\cite{padilla07,schurig06}. The magnetic component of light
couples to both the center section of the electric resonator and the
cut wire, thus generating antiparallel currents resulting in
resonant $\mu(\omega)$ response. The magnetic response can therefore
be tuned independently of the electric resonator by changing the
geometry of the cut wire and the distance between elements. By
tuning each of the resonances it is possible to approximately match
the impedance ($Z=\sqrt{\mu/\epsilon}$) to free space, i.e.
($\epsilon = \mu) \Rightarrow (Z=Z_0$) and minimize the reflectance
at a specific frequency.

\begin{figure}
[ptb]
\begin{center}
\includegraphics[keepaspectratio=true]%
{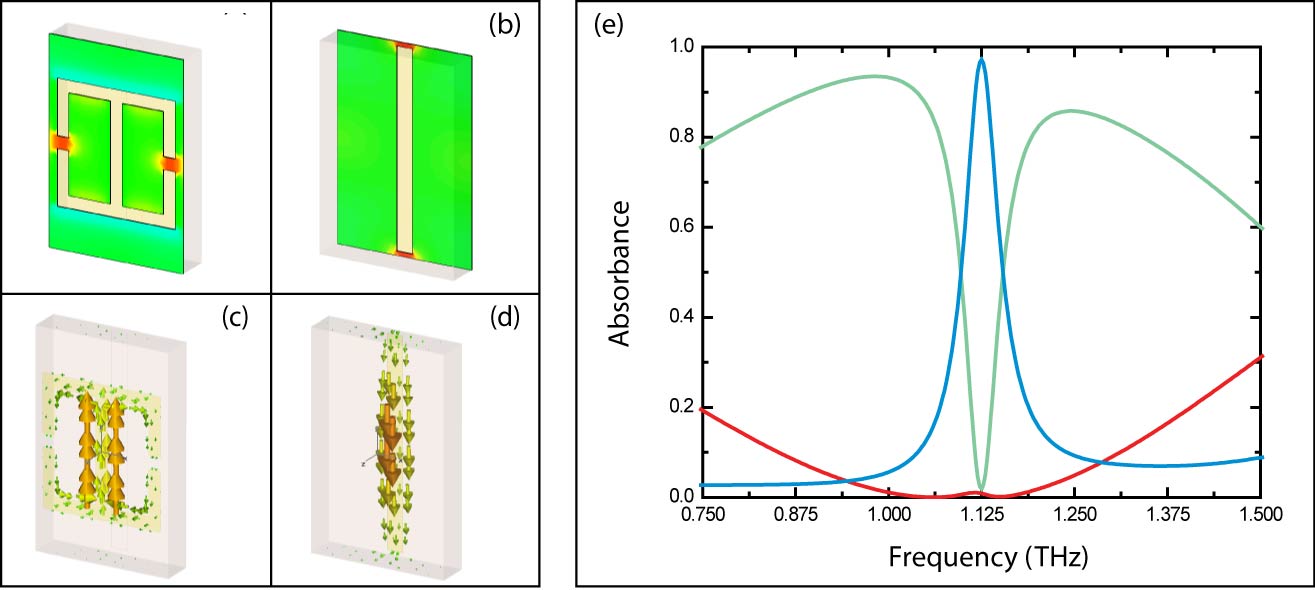}%
\caption{(color) Simulation results for the electric resonator ring
and cut wire. (a) and (b) show the x-component of the electric field
of the electric resonator ring and cut wire at resonance,
respectively; (c) and (d) show the anti-parallel currents driven by
magnetic coupling. (e) The absorptivity (blue) yields a value of
98\% at 1.12 THz. Reflection (green) and Transmission (red) are both
at normal incidence.}
\label{Fig2}%
\end{center}
\end{figure}

Computer simulations were performed using the commercial
finite-difference time domain solver CST Microwave Studio $^{TM}$
2006B and 2008. The metamaterials, depicted in Fig. \ref{Fig1} were
modeled as lossy gold with a conductivity of $\sigma =$
1.0$\times10^7$ S/m. The bottom substrate was modeled as gallium
arsenide with a dielectric constant of 10.75. A 8 $\mu$m thick layer
of dielectric, $\widetilde{\epsilon}$=3.5+i0.02 was used as the
spacer between the two metallic metamaterial elements. We first
investigated the S-parameters of transmission ($\widetilde{S}_{21}$)
and reflection ($\widetilde{S}_{11}$) of a single unit cell with
Perfect Electric (PE) and Perfect Magnetic (PM) boundary conditions
along the $\hat{x}$ and $\hat{y}$ directions, respectively, (see
Fig. \ref{Fig1}). The absorptivity was calculated using the equation
$A=1-|{S}_{21}|^2-|{S}_{11}|^2$. The electric and magnetic fields
were examined at resonance to verify that we were coupling to the
correct resonant mode of each metamaterial element. In Fig.
\ref{Fig2}, the resonant component of the electric field at
resonance is plotted for the electric ring resonator (ERR) (a) and
the split wire (b). The electric field is concentrated strongly in
the gaps of the ring resonator and at the edges of the split wire in
accord with previous results \cite{padilla07,schurig06}. Figure
\ref{Fig2} (c) and (d) show a vector plot of the surface current
density for the ERR and the split wire, respectively. Notice that at
resonance currents are anti-parallel in the two metamaterial
elements, which is the basis of the magnetic response and consistent
with previous results \cite{landy07,dolling07}.

\begin{figure}
[ptb]
\begin{center}
\includegraphics[keepaspectratio=true]%
{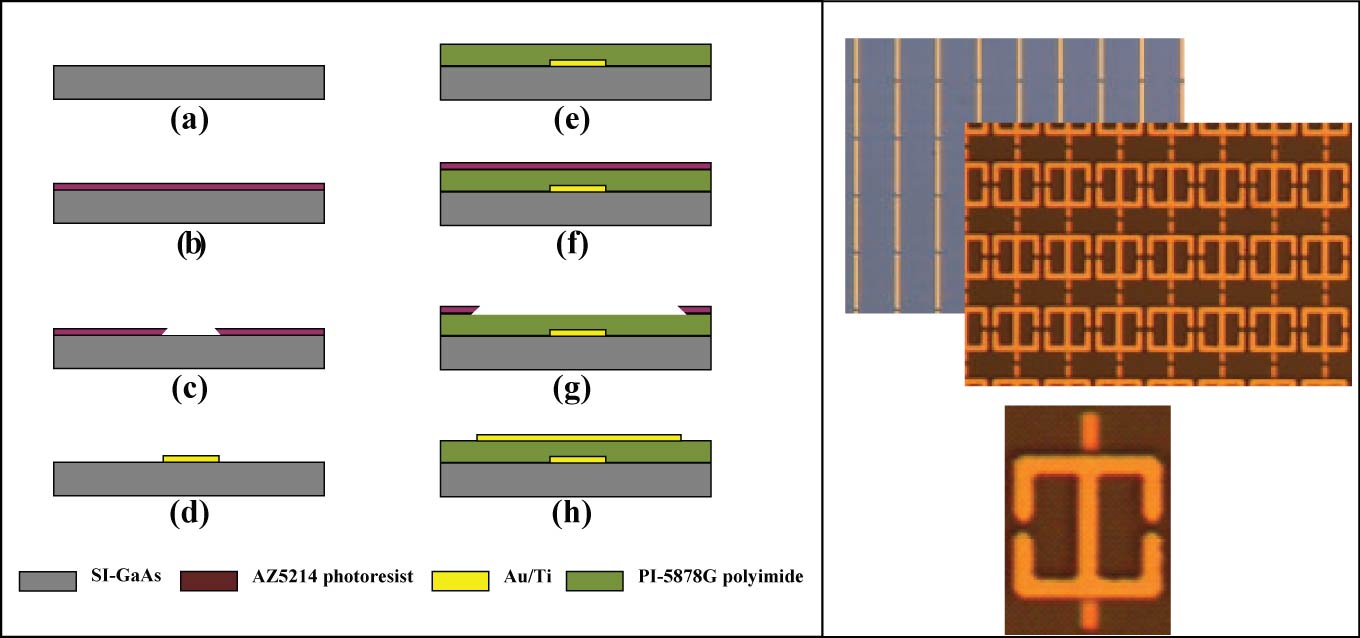}%
\caption{(color) Left panel describes the development process for
fabrication of the terahertz absorber. Right panel shows photographs
of the split wire (top) electric ring resonator and split wire (middle) and an
individual unit cell of the terahertz absorber (bottom). }
\label{Fig4}%
\end{center}
\end{figure}

By changing the electric and magnetic resonances individually, we
were able to create a condition such that the material was at an
impedance near the free space value in a region of very low
T($\omega$). The simulated transmission is relatively low across the
entire range shown in Fig. \ref{Fig2} (e), whereas the reflectivity
is relatively high except near the resonance at 1.12 THz where it
drops to a value of 2$\%$. Near unity absorbtion is theoretically
possible and here we achieve a simulated value of 98$\%$ at 1.12
THz, as shown in Fig. \ref{Fig2}(e). It should be noted that at THz
frequencies the magnetic response of natural and metamaterials is
significantly weaker than the electric. Thus, matching the exact
form of the $\epsilon(\omega)$ and $\mu(\omega)$ resonances becomes
increasingly difficult at these and higher frequencies. Further,
from a viewpoint of spectrally selective thermal imaging, it is
desirable to have a narrow-band absorber. Thus we strive for two
requirements of our metamaterials at our target frequency, i.e.
$\epsilon_1=\mu_1$ and $\epsilon_2$,$\mu_2\sim0$. With these two
goals we can achieve a significant and narrow-band $A(\omega)$, but
tolerance limits associated with microfabrication can reduce the
absorptivity from unity.

\begin{figure}
[ptb]
\begin{center}
\includegraphics[keepaspectratio=true]%
{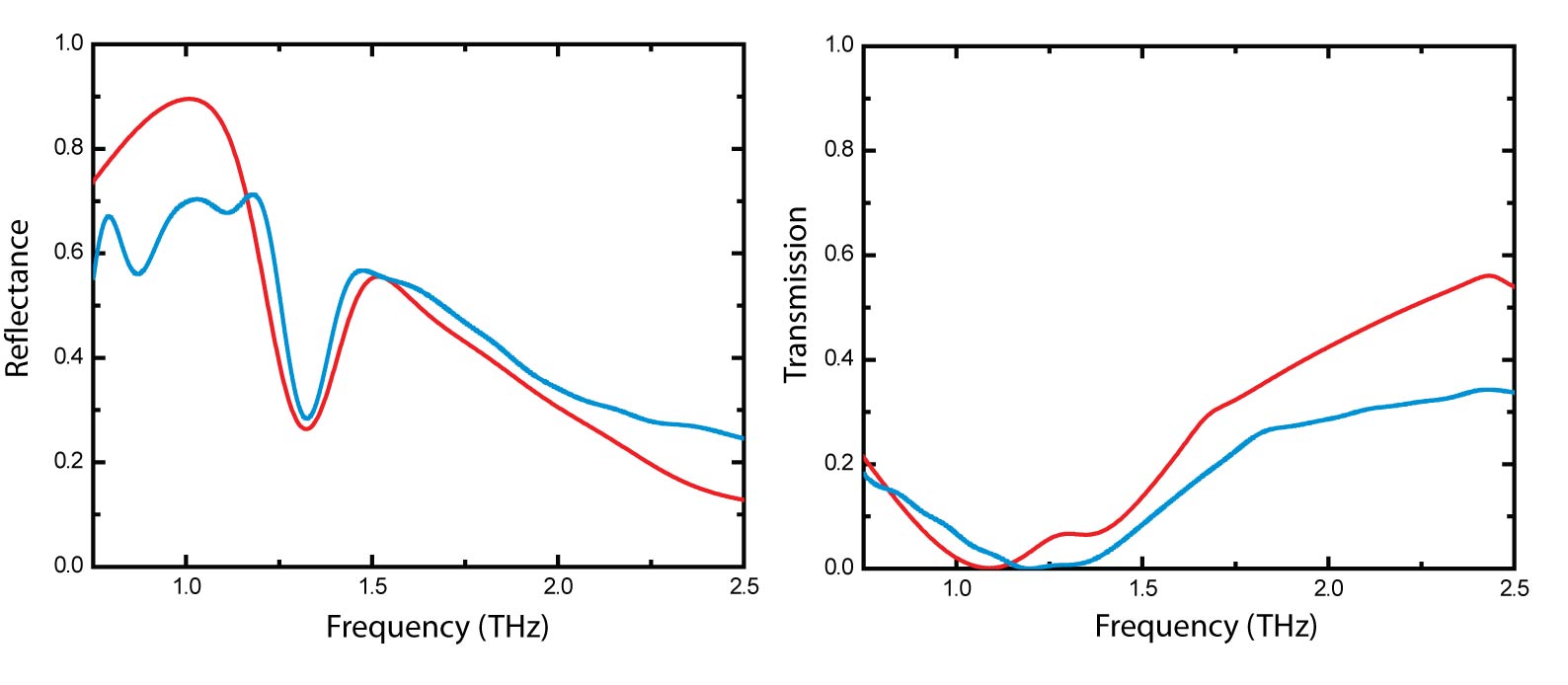}%
\caption{(color) Experimental results showing the transmission
intensity and reflection intensity. The blue lines are experiment
and the red line the simulations. The reflectance measurement was
performed at $30^{\circ}$ off-normal. The transmission measurement was
performed at normal incidence.}
\label{Fig5}%
\end{center}
\end{figure}

We fabricated the metamaterial shown in Fig. \ref{Fig1} using a
surface micromachining process, as shown in Fig. \ref{Fig4}. A
semi-insulating GaAs wafer was chosen because it is highly
transmissive at THz frequencies. AZ5214e image reversal photoresist
was spin-coated and patterned using standard photolithography. A 200
nm-thick Au/Ti film was E-beam evaporated to create the cut wire on
the bottom layer. Lift off of the photoresist was achieved by
rinsing in acetone for several minutes. The liquid polyimide, HD
Microsystems $^{TM}$ PI-5878G, was spin-coated at 2,000 rpm on the
GaAs wafer to form an insulating spacer with a thickness of 8
$\mu$m, and cured for five hours in an oven at $275^\circ$C in a
nitrogen environment after the soft bake at $110^\circ$C for 6
minutes on a hot plate. AZ5214e image reversal photoresist was
spin-coated, aligned, and patterned using standard photolithography.
Another 200 nm-thick Au/Ti was E-beam evaporated as the material of
the electric resonant ring on the top layer and then lifted off.
Microscopic images of the as-fabricated samples are shown in Fig.
\ref{Fig4} (right).

We experimentally verified the behavior of the absorber by measuring
the transmission and reflectance of a large (1 cm$\times$1 cm)
planar array. We used an evacuated Fourier transform infrared (FTIR)
spectrometer in the range from 300 GHz - 3 THz (10 - 100 cm$^{-1}$)
with 15 GHz (0.5 cm$^{-1}$) spectral resolution. For transmission
measurements the sample was mounted in the FTIR at normal incidence
with the electric field perpendicular to the gap of ERR, as depicted
in Fig. \ref{Fig1}(a). Reflection was performed at an angle of
$30^\circ$ due to experimental limitations. The blue curves in Fig.
\ref{Fig5}(a) and (b) show, respectively, the measured reflectivity
and transmission. Measured R($\omega$) and T($\omega$) differ
significantly from that simulated as shown in Fig. \ref{Fig2}(e).
However, it should be noted that values used in simulation for the
polyimide spacing layer were estimated based on published values at
lower GHz frequencies. Further, the thickness of the polyimide layer
was measured to be closer to 6$\mu$m rather than the 8$\mu$m used in
simulations. Taking both of these factors into account, we are able
to match measurements using an experimentally determined value for
polyimide of $\epsilon=2.5+i0.2$. The red curves are the
corresponding simulations which are in good agreement with
experiment. The simulated reflectance matches reasonably well near
the resonance with slight deviations at lower and higher
frequencies. The simulated transmission also agrees well with
experiment, particularly in the vicinity of the resonance. Further,
the simulated T($\omega$) reproduces the same qualitative features
as experiment, including a distinct kink near 1.75 THz.

\begin{figure}
[ptb]
\begin{center}
\includegraphics[keepaspectratio=true]%
{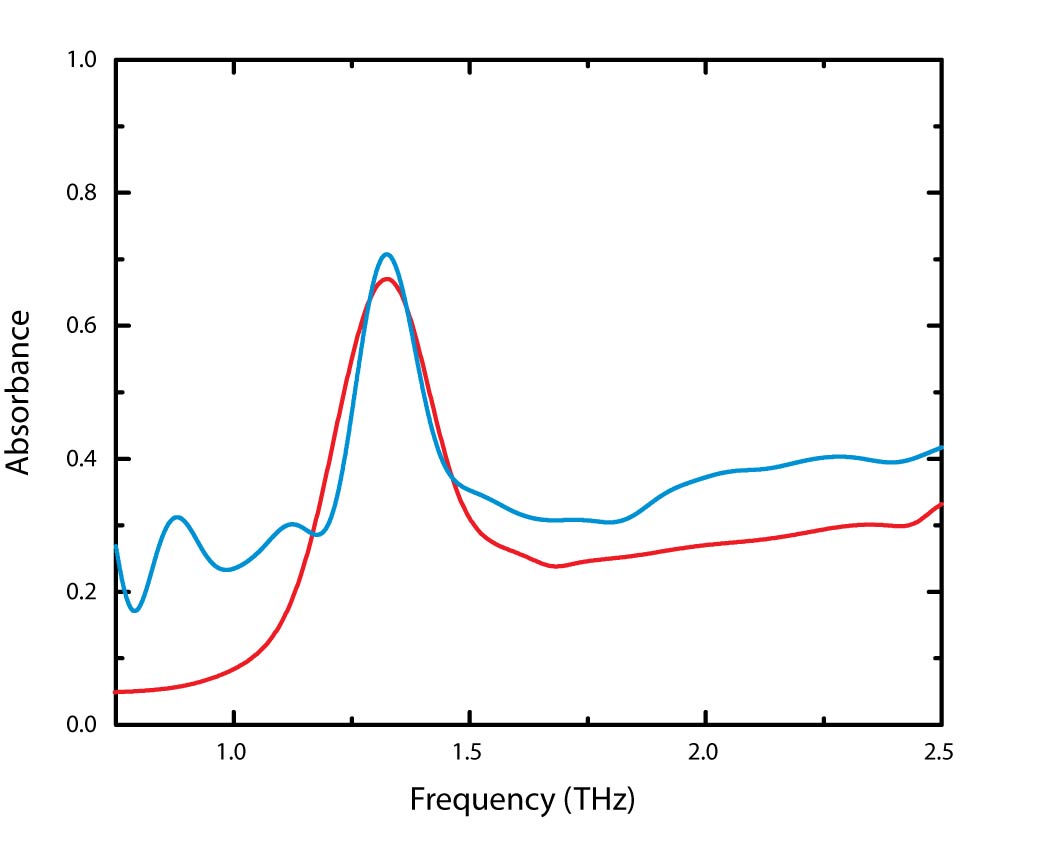}%
\caption{(color) Experimental results showing absorptivity.
Experimental results are in blue and simulation is in red. The
experimental absorptivity reaches a maximum value of 70\% at 1.3
THz. The simulated absorptivity reaches a value of 68\% at the same
frequency.}
\label{Fig6}%
\end{center}
\end{figure}

From experimental data presented in Fig. \ref{Fig5}, the
corresponding absorptivity is determined as shown in Fig.
\ref{Fig6}. These results demonstrate that the as-fabricated MMs
have a strong resonance at around 1.3 THz and a high absorptivity of
approximately 70\%. Simulations show good agreement with experiment
(red). The simulated absorptivity at resonance matches very well
with the measured value. The off-resonance absorptivity is higher in
experiment than in simulation due to differences in the experimental
and simulated S-parameters. The response of absorber could be
further improved through refinement and optimization of the
fabrication process.

\begin{figure}
[ptb]
\begin{center}
\includegraphics[keepaspectratio=true]%
{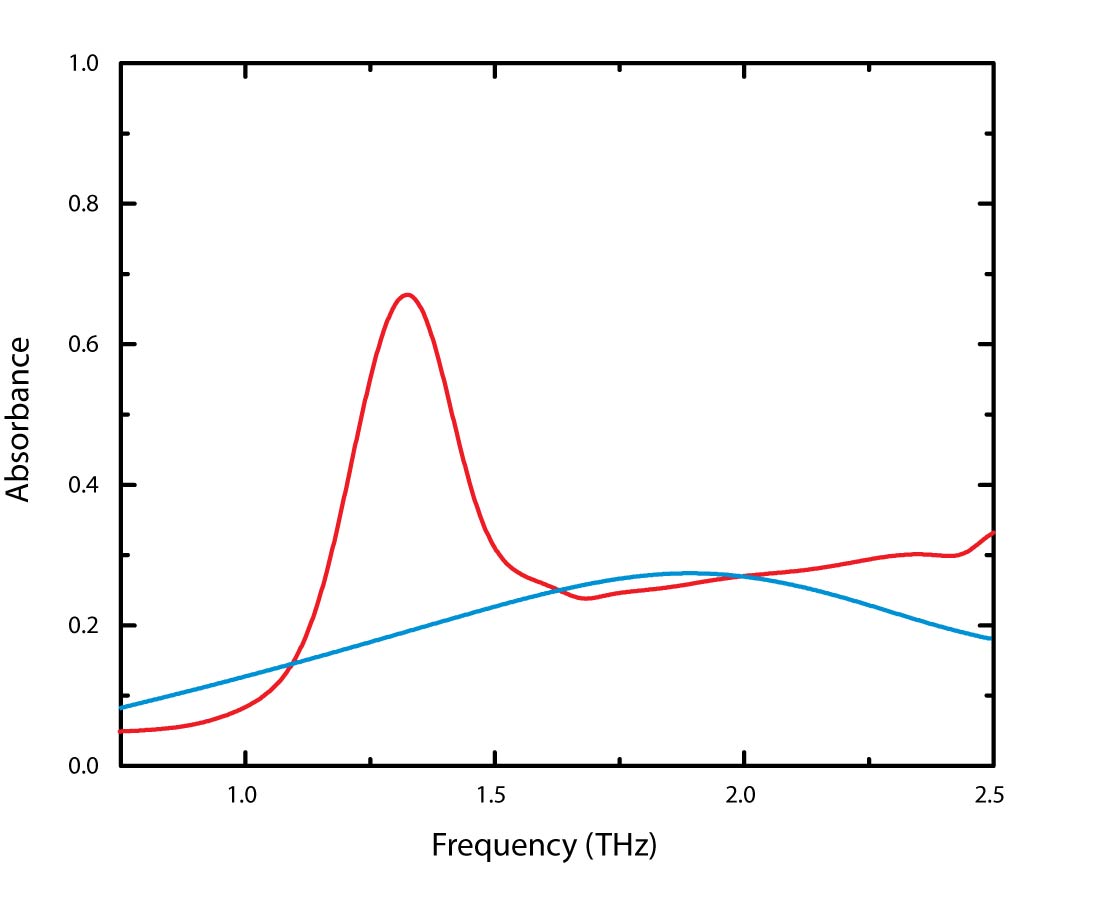}%
\caption{(color) Simulation results comparing absorptivity for both
polarizations. When the electric field is polarized parallel to the
center stalk of the ERR (red) absorption reaches 70\%. In the
opposite polarization, the absorption only reaches 27\%}
\label{Fig7}%
\end{center}
\end{figure}

The absorber presented in this work absorbs strongly for light
polarized along the $\hat{x}$ direction, as shown in Fig.
\ref{Fig1}, but poor for $\hat{y}$ polarized light, as shown in Fig.
\ref{Fig7}. In this polarization, the electric field is
perpendicular to the center stalk of ERR, so an electric response
cannot be driven. Similarly, there are no parallel wires for the
magnetic field to develop a flux through and thus no net magnetic
response. Such a polarization-sensitive device is desirable for both
mm-wave and THz imaging as reflections from metallic objects often
saturate the imager, thus significantly degrading it performance, in
a problem known as ``glint". Additionally, polarization-sensitive
detection has been shown to aid in discrimination of objects in a
scene \cite{cremer01,sadjadi03}. However, we have designed higher
symmetry metamaterials, (similar to those presented here), which
achieved similar absorptive behavior but are polarization
insensitive. Furthermore, a mechanically flexible high absorber is
desirable so that it can conform to a non-planar surface, which will
expand potential applications significantly including, for example,
electromagnetic cloaking \cite{pendry06,schurig06a}.

The performance of a THz radiation detector depends on the
efficiency of converting radiation energy to an output signal.
Therefore, maximizing the THz radiation absorption efficiency is
integral to the development of a functional THz detector/imager. It
is difficult to find strongly absorbing materials at THz frequencies
that are compatible with standard photolithography. Thus, a
potential application of these metamaterial structures is as the
absorbing elements in thermal detectors. A strong absorption
coefficient is also necessary to have a small thermal mass. This is
important for optimizing the temporal response of thermal detectors.
The metamaterial presented here has a 6 micron thick film and 70\%
absorptivity, which yields an absorption coefficient of 2000
cm$^{-1}$. Further optimization could increase this value by nearly
an order of magnitude.

Many micro thermal detectors (typically based on bolometric
detection and appropriately termed microbolometers) utilize several
materials as the sensing element such as VOx~\cite{white98},
poly-Si-Ge~\cite{wauters97}, YBCO~\cite{almasri01}, or metal
resistors such as, titanium~\cite{lee99} and
niobium~\cite{baorino99}. Some of these materials are not fully
compatible with microfabrication processing. For those that are,
however, it is difficult and thus expensive to prepare or deposit a
high quality film. Additionally, most of these materials show
broad-band absorption. This limits potential applications, such as
spectroscopic detection of explosive materials, which show unique
responses at varied frequencies~\cite{barber05}. The narrowband
absorptivity of metamaterials presented here, on the other hand,
enable spectrally selective detection. Furthermore, MMs are
geometrically scalable and have been demonstrated over many decades
of frequency. Thus, our results are not limited to terahertz
frequencies and may be used over much of the electromagnetic
spectrum. Another salient feature of the design presented here is
that it may be combined with semiconducting materials or
ferroelectrics to enable optically or electrically tunable frequency
agile terahertz metamaterials. This would further permit a
hyperspectral metamaterial focal plane array imager able to imaging
over a relatively large band \cite{chen08}. Planar metamaterial
absorbers consisting of different unit cells with distinct resonance
frequencies may permit ``multi-color" imaging.

In summary, we have demonstrated that the electromagnetic response
of metamaterials can be tailored by manipulating the geometries of
electric and magnetic resonators individually to create a highly
selective absorber over a narrow band at THz frequencies. The
successful demonstration of the high absorber holds great promise
for future applications which includes metamaterial-based structures
for creating a narrow-band, low thermal mass absorber as required
for thermal sensing applications.

Acknowledgement: NIL, CMB, RDA and WJP acknowledge support from the
Los Alamos National Laboratory LDRD program. This project has been
supported in part by the DOD/Army Research Laboratory through grant
W911NF-06-2-0040, DOE Los Alamos National Laboratory subcontract
50332-001-07 and DOE Los Alamos National Laboratory subcontract
50335-001-07. The authors would like to thank the Photonics Center
at Boston University for technical support throughout the course of
this research.

\end{document}